\documentclass[11pt, reqno]{article}
\pdfoutput=1

\usepackage{graphicx}
\graphicspath{ {./figures/} }
\usepackage{subfig}
\usepackage{jheppub}
\usepackage{amssymb}
\usepackage{amsmath}
\usepackage[usenames,dvipsnames]{xcolor}
\usepackage{epsfig}
\usepackage{dcolumn}
\usepackage{tikz}
\usetikzlibrary{shapes.geometric, arrows}
\usepackage{upgreek}
\usepackage{setspace}
\usepackage{enumitem}
\usepackage{array,multirow,bigdelim,arydshln}
\usepackage{appendix}
\usepackage{xparse}
\usepackage[utf8]{inputenc}
\usepackage{hyperref}
\hypersetup{
	colorlinks,
	urlcolor=Maroon,
	linkcolor=Maroon,
	citecolor=Maroon
}

\usepackage{amsthm}

\theoremstyle{definition}

\usepackage{mathtools}

\usepackage{float}
\restylefloat{table}

\NewDocumentCommand{\binomial}{omm}
 {%
  \genfrac(){0pt}{}{#2}{#3}%
  \IfValueT{#1}{_{\!#1}}%
 }
\NewDocumentCommand{\eulerian}{omm}
 {%
  \genfrac<>{0pt}{}{#2}{#3}%
  \IfValueT{#1}{_{\!#1}}%
 }

\def \s {\sigma}

\usepackage{underscore}

\usepackage{latexsym}
\usepackage{tikz}

\newcommand{\lgg}{\langle}
\newcommand{\rgg}{\rangle}

\title{Computing NMHV Gravity Amplitudes at Infinity}

\author[a,b]{Dawit Belayneh,}\emailAdd{dbelayneh@pitp.ca}
\author[a]{Freddy Cachazo,}\emailAdd{fcachazo@pitp.ca}
\author[a]{and Pablo Leon}\emailAdd{pleon@pitp.ca}

\affiliation[a]{Perimeter Institute for Theoretical Physics, Waterloo, ON N2L 2Y5, Canada}
\affiliation[b]{Department of Physics \& Astronomy, University of Waterloo, Waterloo, ON N2L 3G1, Canada}

\abstract{In this note we show how the solutions to the scattering equations in the NMHV sector fully decompose into subsectors in the $z\to \infty$ limit of a Risager deformation. Each subsector is characterized by the punctures that coalesce in the limit. This naturally decomposes the $E(n-3,1)$ solutions into sets characterized by partitions of $n-3$ elements so that exactly one subset has more than one element. We present analytic expressions for the leading order of the solutions in an expansion around infinite $z$ for any $n$. We also give a simple algorithm for numerically computing arbitrarily high orders in the same expansion. As a consequence, one has the ability to compute Yang-Mills and gravity amplitudes purely from this expansion around infinity. Moreover, we present a new analytic computation of the residue at infinity of the $n=12$ NMHV tree-level gravity amplitude which agrees with the results of Conde and Rajabi. In fact, we present the analytic form of the leading order in $1/z$ of the Cachazo-Skinner-Mason/CHY formula for graviton amplitudes for each subsector and to all multiplicity. As a byproduct of the all-order algorithm,  one has access to the numerical value of the residue at infinity for any $n$ and hence to the corrected CSW (or MHV) expansion for NMHV gravity amplitudes.}

\begin{document}
\maketitle
\addtocontents{toc}{\protect\setcounter{tocdepth}{1}}
\def \tr {\nonumber\\}
\def \nn {\nonumber}
\def \la {|}
\def \ra {|}
\def \dd {\Theta}
\def\hset{\texttt{h}}
\def\gset{\texttt{g}}
\def\sset{\texttt{s}}
\def\A {\textsf{A}}
\def\B {\textsf{B}}
\def\C {\textsf{C}}
\def\D {\textsf{D}}
\def\E {\textsf{E}}
\def\F {\textsf{F}}
\def\G {\textsf{G}}
\def\I {\textsf{I}}
\def\J {\textsf{J}}
\def\H {\textsf{H}}
\def \be {\begin{equation}}
\def \ee {\end{equation}}
\def \ba {\begin{eqnarray}}
\def \ea {\end{eqnarray}}
\def \k {\kappa}
\def \h {\hbar}
\def \r {\rho}
\def \l {\lambda}
\def \be {\begin{equation}}
\def \en {\end{equation}}
\def \bes {\begin{eqnarray}}
\def \ens {\end{eqnarray}}
\def \red {\color{Maroon}}
\def \pt {{\rm PT}}
\def \s {\textsf{s}}
\def \t {\textsf{t}}
\def \C {\textsf{C}}
\def \tp {||}
\def \p {x}
\def \x {z}
\def \V {\textsf{V}}
\def \ls {{\rm LS}}
\def \ma {\Upsilon}
\def \SL {{\rm SL}}
\def \GL {{\rm GL}}
\def \w {\omega}
\def \e {\epsilon}
\def \a {\alpha}
\def \g {\gamma}
\def \b {\beta}
\def \ort {\textsf{O}}

\numberwithin{equation}{section}

\section{Introduction}

The analytic structure of tree-level scattering amplitudes of massless particles can be used to compute them in a variety of ways \cite{Elvang:2013cua}. Two examples are the BCFW recursion relations \cite{Britto:2004ap,Britto:2005fq}, which use a one-complex dimensional deformation, and the Witten-RSV \cite{Witten:2003nn,Roiban:2004yf} or CHY formulations \cite{Cachazo:2013hca}, which encode the physical singularities in the moduli space of punctured Riemann spheres, ${\cal M}_{0,n}$. In this work, we show that these two techniques are connected in an unexpected way. 

We start with a generalization of the BCFW construction known as the Risager deformation \cite{Risager:2005vk}. This is a one-complex dimensional deformation of a four-dimensional scattering amplitude designed to reproduce the CSW (or MHV) expansion of Yang-Mills amplitudes \cite{Cachazo:2004kj} following the same steps as in a BCFW construction. 

The first non-trivial use of a Risager deformation is in the NMHV sector of Yang-Mills amplitudes, where it directly leads to the CSW expansion \cite{Risager:2005vk}. The same procedure was applied to gravity amplitudes and its validity checked for $n<8$ gravitons \cite{Bjerrum-Bohr:2005xoa}. Unfortunately, the Risager deformation fails to produce a CSW-like expansion for gravity amplitudes for twelve or more gravitons due to the presence of poles at infinity \cite{Bianchi:2008pu}.   

The next ingredient in this work is the scattering equations used in the CHY formulation of scattering amplitudes \cite{Fairlie:1972zz,RobertsThesis,Fairlie:2008dg,Cachazo:2013gna,Cachazo:2013hca,Cachazo:2013iea}. In general dimensions, the scattering equations have $(n-3)!$ solutions and for generic kinematics the solutions define generic points on ${\cal M}_{0,n}$. In four dimensions, the solutions split into sectors traditionally labeled by an integer $k$, with $2\leq k\leq n-2$. The solutions in each sector coincide with those obtained using the Witten-RSV equations in the ${\rm N}^{k-2}{\rm MHV}$ sector. 

The number of solutions in the $k$ sector is $E(n-3,k-2)$  \cite{Spradlin:2009qr,Cachazo:2013iaa}, where $E(a,b)$ is an Eulerian number and counts the number of permutations of the $a$ elements with $b$ ascents \cite{oeis8292}. For example, $E(n-3,0)=1$ and $E(n-3,1) = 2^{n-3}-n+2$.  

In this work we concentrate on the $k=3$ sector, i.e., the ${\rm NMHV}$ sector, and study how the $E(n-3,1) = 2^{n-3}-n+2$ solutions behave under the Risager deformation of three particle's momenta. The one-complex parameter, $z$, deforms the spinors as follows \cite{Risager:2005vk},
\be
\tilde\lambda_a(z) = \tilde\lambda_a + z \langle b~c\rangle \tilde\mu , 
\ee 
where $(a,b,c)$ is a cyclic permutation of $(1,2,3)$ and $\tilde\mu$ is a fixed generic reference spinor. All other spinors are left undeformed. 

We apply this deformation to the invariants $s_{ab}$ in the scattering equations and study the limit $z\to \infty$. 

Surprisingly, we find that all solutions in the {\rm NMHV} sector separate into subsectors with a single solution in each. Each subsector is characterized by the punctures that coalesce. If $u_1,u_2,u_3$ are gauge fixed, then the sectors correspond to all possible ways of partitioning the set $\{ 4,5,\ldots, n\}$ into sets such that exactly one set has more than one element \cite{oeis295}. This turns out to be one of the many interpretations of $E(n-3,1)$. We find analytic expressions for the leading order of all $2^{n-3}-n+2$ solutions in an expansion in $1/z$ for all $n$. 

We also provide a simple algorithm for computing the $1/z$ expansion to arbitrarily high orders and find that the radius of convergence is in general determined by the branch points of the solutions.

Finally, we present applications of the expansions. The first is an analytical computation of the residue at infinity of the $n=12$ graviton NMHV amplitude under the Risager deformation which agrees with the result obtained by Conde and Rajabi \cite{Conde:2012ik} but which manifests the symmetries of the problem and hints at a physical interpretation. Moreover, we give the explicit form of the leading order contributions to the amplitude coming from each subsector and find that the more points coalesce in the large $z$ limit, the higher the order in $1/z$. This means that the overall leading order is controlled by the subsectors with exactly two points that coalesce. 

The second application is a (semi) numerical technique for computing NMHV scattering amplitudes in Yang-Mills and in gravity which trades poles at finite locations in the deformation parameter space for poles at infinity. Since we have access to large orders in the expansion of the corresponding scattering equations solutions, it is possible to evaluate the residues at infinity. 

In these applications, we see that, at least for the NMHV, the limit $z\to \infty$ is well under control and it is in fact a convenient place to perform computations.

This paper is organized as follows. In section \ref{reviewForm} we give a brief review of standard constructions used in this work. In section \ref{chSects} we consider the behavior of NMHV solutions to the scattering equations in the large $z$ limit and give the explicit form of the leading order term in an expansion in $1/z$. In section \ref{chAlg} we give the details of an algorithm for computing the $1/z$ expansion of NMHV solutions to any order and we consider the radius of convergence of the $1/z$ series expansion, including a discussion of the Mercer-Roberts estimator. In section \ref{chNMHV}, we present the first application of the expansions by computing the leading order of NMHV gravity amplitudes in a $1/z$ expansion for large $z$ under the Risager deformation. Here is where we compute the full $n=12$ residue at infinity. In section \ref{chCInf}, we propose a second application; using the expansion to numerically compute NMHV amplitudes at infinity. Section \ref{chDis} contains discussions of our results with some future directions. We end with several appendices that detail some of the results used in this work.

\section{Review of Risager's Deformation, Witten-RSV, CSM, and CHY formulations}\label{reviewForm}

This section reviews standard constructions and it is included to set the notation and for the reader's convenience.

\subsection{Risager's Deformation}

The kinematic data for the scattering of $n$ massless particles in four dimensions can be specified in terms of spinors 
\be
\{ \{ \lambda_1,\tilde\lambda_1\}, \ldots ,\{ \lambda_n,\tilde\lambda_n\}\}
\ee 
satisfying momentum conservation
\be
\sum_{a=1}^n \lambda_a\tilde\lambda_a = 0.
\ee
The Risager deformation tailored to NMHV amplitudes deforms the anti-holomorphic spinors of three particles as follows:
\be\label{risagerDeformation}
\tilde\lambda_a(z) = \tilde\lambda_a + z \langle b~c\rangle \tilde\mu , 
\ee 
where $(a,b,c)$ is a cyclic permutation of $(1,2,3)$ and $\tilde\mu$ is a fixed generic reference spinor. All other spinors are left undeformed.

When this deformation is applied to NMHV amplitudes of gluons, amplitudes with helicities $1^-,2^-,3^-,4^+,\ldots , n^+$ only have poles at factorization channels where the two resulting sub-amplitudes are both MHV amplitudes. This leads to the derivation of the CSW rules in the NMHV sector.

\subsection{Witten-RSV and CSM Formulations} \label{chWRSV}

Formulas for tree amplitudes of gluons in terms of rational maps of various degrees were proposed by Witten in \cite{Witten:2003nn} and developed by Roiban, Spradlin, and Volovich (RSV) in \cite{Roiban:2004yf}. Based on the same rational maps, formulas for amplitudes of gravitons were presented by Cachazo, Skinner \cite{Cachazo:2012kg} and with Mason in \cite{Cachazo:2012pz} (CSM). Here we review rational maps for any holomorphic degree, $d$, but keeping in mind that the NMHV sector corresponds to $d=2$.

Consider a $\mathbb{CP}^1$ with inhomogenous coordinate $u$. The following function
\be
    \lambda(u)  :=  \rho_0+\rho_1 u+\ldots +\rho_d u^d,
\ee
where $\rho_m$ are spinors, defines a rational map of degree $d$ onto $\mathbb{CP}^1$ provided the equations $\lambda_1(u)=0$ and $\lambda_2(u)=0$ do not have simultaneous solutions. 

The Witten-RSV formulation of gluon amplitudes associates a puncture $u_a$ to each particle via the equations
\begin{align}
   & \lambda_a = t_a \lambda(u_a) \quad {\rm for}\quad a\in \{ 1,2,\ldots ,n\}, \\
    & \sum_{a=1}^n t_a\tilde\lambda_a u_a^m = 0 \quad {\rm for}\quad m\in \{ 0,1,\ldots ,d\} . 
\end{align} 
Here $t_a$ are auxiliary variables introduced to explicitly impose the projective invariant equation $\langle \lambda_a , \lambda(u_a)\rangle = 0$. These equations have a $GL(2,\mathbb{C})$ redundancy that can be fixed by choosing the values of, e.g., $\{ u_1,u_2,u_3,t_1\}$. 

Amplitudes of gluons/gravitons with $k$ negative helicity and $n-k$ positive helicity particles are computed as integrals localized to the solutions of the $d=k-1$ equations. In the case of $\mathcal{N}=4$ Super Yang-Mills this is the Witten-RSV integral formula given by

\be \label{eqWRSV}
A_n^{(k=d+1)} = \int d \mathcal{M}_{n,d} \>\> \prod_{a=1}^n \delta^2(\lambda_a - t_a \sum_{m=0}^d \rho_m u_a^m) \>\> \prod_{m=0}^d \delta^2(\sum_{a=1}^n t_a u_a^m \tilde \lambda_a)\delta^4(\sum_{a=1}^n t_a u_a^m \eta_a),
\ee
where $\eta_a$ is a four-component Grassmann variable and the measure is given by:
\be 
d \mathcal{M}_{n,d} = \frac{1}{\text{Vol}(GL(2, \mathbb{C}))}\prod_{m=0}^d d^{2}\rho_m \prod_{a=1}^n \frac{dt_a \>\> du_a}{t_a(u_a - u_{a+1})}.
\ee
At tree level, supersymmetric gluon amplitudes coincide with that of regular Yang-Mills theory. As a result, we may use the integral \eqref{eqWRSV} as a generating function for tree level amplitudes with any configuration of $k$ negative helicity gluons. For example, if gluons $1,2,\ldots ,k$ have negative helicity, then the Grassmann integrals give rise to a factor of $(|12\cdots k|t_1t_2\dots t_k)^4$, where $|12\cdots k|$ is the Vandermonde determinant of $u_1,u_2,\ldots ,u_k$.

Here we point out that the integral \eqref{eqWRSV} contains a momentum conserving delta function $\delta^4(\sum_a\lambda_a \tilde \lambda_a)$ which must be extracted before it is evaluated. Once this has been accomplished and we gauge fix the $GL(2,\mathbb{C})$ redundancy we will have $2(n+d-1)$ integrations over $2(n+d-1)$ delta functions. As a result, the integral localizes to a discrete sum over the solutions of the Witten-RSV equations.   

Similarly, the CSM formula for $\text{N}^{k-2}\text{MHV}$ amplitudes in $\mathcal{N}=8$ supergravity is given by 
\be 
M_n^{(k=d+1)} = \int d \mathcal{M}_{n,d} \>\> |\Phi|'|\tilde{\Phi}|' \prod_{a=1}^n \delta^2(\lambda_a - t_a \sum_{m=0}^d \rho_m u_a^m) \>\> \prod_{m=0}^d \delta^2(\sum_{a=1}^n t_a u_a^m \tilde \lambda_a)\delta^8(\sum_{a=1}^n t_a u_a^m \eta_a),
\ee
where $\eta_a$ is the corresponding eight-component Grassmann variable and the measure is given by
\be 
d \mathcal{M}_{n,d} = \frac{1}{\text{Vol}(GL(2, \mathbb{C}))}\prod_{m=0}^d d^{2}\rho_m \prod_{a=1}^n \frac{dt_a \>\> du_a}{t^3_a}.
\ee
The reduced determinants $|\Phi|'$ and $|\tilde{\Phi}|'$ are a generalization of Hodges’ formulation of MHV amplitudes and are defined as follows: Let $\Phi$ be the symmetric $n\times n$ matrix with elements 
\bes 
    \Phi_{ij} &=& \frac{\langle i ~ j \rangle}{u_i-u_j}\frac{1}{t_it_j} \quad \mbox{for} \quad i\neq j, \\
    \Phi_{ii} &=& - \sum_{j\neq i} \Phi_{ij}\prod_{r=0}^{n-d-2} \left(\frac{u_j-v_{r}}{u_i-v_r} \right) \frac{\prod_{k \neq i} (u_i-u_{k})}{\prod_{l \neq j} (u_j-u_{k})}
\ens
where $v_{r}$ are reference points on $\mathbb{CP}^1$ and $\tilde{\Phi}$ the $n\times n$ matrix with elements
\bes
    \tilde{\Phi}_{ij} &=&  \frac{[ i ~ j]}{u_i-u_j}t_it_j \quad \mbox{for} \quad i\neq j, \\
    \tilde{\Phi}_{ii} &=& - \sum_{j\neq i} \tilde{\Phi}_{ij}\prod_{a=0}^{d} \frac{u_j-v_a}{u_i-v_a}.
\ens
Then $|\Phi|'$ is defined as
\be
    |\Phi|' := \frac{\det(\Phi_{red})}{|r_1...r_d||c_1...c_d|},
\ee
where $\Phi_{red}$ is a non-singular matrix obtained by removing any $n-d$ rows and columns of $\Phi$. The labels $r_1...r_d$ and $c_1...c_d$ correspond to the rows and columns that remain in $\Phi_{red}$, respectively.  
Similarly, we can define a non-singular matrix, $\tilde{\Phi}_{red}$, removing any $d+2$ rows and columns of $\tilde{\Phi}$ to define $|\tilde{\Phi}|'$ as 

\be
    |\tilde{\Phi}|' := \frac{\det(\tilde \Phi_{red})}{|\tilde{r}_1...\tilde{r}_{d+2}||\tilde{c}_1...\tilde{c}_{d+2}|}.
\ee
In this case, $\tilde{r}_1...\tilde{r}_d$ and $\tilde{c}_1...\tilde{c}_d$ are the labels of the deleted rows and columns of $\tilde{\Phi}$, respectively.

\subsection{CHY Formulation}

The Cachazo-He-Yuan (CHY) formulation is built on the observations that there is a deep connection between the moduli space of punctured Riemann spheres, ${\cal M}_{0,n}$, and the singularities of scattering amplitudes in arbitrary space-time dimensions. The connection is achieved using the function on ${\cal M}_{0,n}$ defined by
\be\label{potentialCHY}
{\cal S} =\sum_{a<b}s_{ab}\log |u_a-u_b|,
\ee 
where $s_{ab}=(k_a+k_b)^2$. The scattering equations are obtained as the conditions for finding the critical points of ${\cal S}$. More explicitly,
\be\label{rationalCHY} 
\frac{\partial {\cal S}}{\partial u_a} = \sum_{b\neq a}\frac{s_{ab}}{u_a-u_b} = 0.
\ee 
Once again, these equations are invariant under a $SL(2,\mathbb{C})$ action which can be used to fix the values of $u_1,u_2$ and $u_3$. 

We postpone the presentation of the integrands for the CHY formulation to the sections where they are needed.    

\section{NMHV Solutions at Infinity Under Risager's Deformation} \label{chSects}

The Witten-RSV equations for $d=2$ and $n$ particles are a complicated set of polynomial equations with $E(n-3,1) = 2^{n-3}-n+2$ solutions. The only known analytic solution for generic kinematics is for $n=5$, in which case the NMHV sector is equivalent to the $\overline{\rm MHV}$ sector. 

In this section we consider the $d=2$ equations for kinematics under the Risager deformation and explore the behavior of the solutions as $z\to \infty$.

Perhaps surprisingly, it turns out to be easier to start with the CHY scattering equations. In general, the scattering equations have $(n-3)!$ solutions. When the kinematic invariants can be written as $s_{ab}=\langle a~b\rangle [a~b]$, the solutions separate into sectors labeled by $k\in \{ 2,\ldots ,n-2\}$. The solutions for $u_a$ in the $k=3$ sector also solve the $d=2$ (or NMHV) Witten-RSV equations.

Noting that under the Risager deformation the only $z$-dependent kinematic invariants  are $s_{1a}(z), s_{2a}(z), s_{3a}(z)$ for $a\in \{ 1,2,\ldots ,n\}$, the equations \eqref{rationalCHY} can be used to extract the leading $z$ term of $u_a(z)$ as $z\to \infty$.

Furthermore, one finds that the $2^{n-3}-n+2$ solutions in the NMHV sector further separate into subsectors. If $u_1$, $u_2$ and $u_3$ are gauge fixed, then the subsectors are conveniently labeled by subsets $J\subset \{4,5,\ldots ,n\}$ with $|J|>1$. It is easy to see that there are exactly $2^{n-3}-n+2$ such subsets. 

Without loss of generality, consider $J_m :=\{4,5,\ldots m\}$, with $m>4$. Other subsets can be obtained by permutation of labels. 

In the $J_m$ sector, there is a unique solution. The large $z$ expansion using the gauge fixing $u_1=0, u_2=\infty$ and $u_3=1$ is of the form
\be
u_a = x_{a,0}+x_{a,1}/z+x_{a,2}/z^2+\ldots 
\ee 
with 
\be\label{leadingOrder}
x_{a,0} = \frac{\langle 1~I_m\rangle\langle 2~3\rangle }{\langle 1~3\rangle\langle 2~I_m\rangle} ~~ {\rm for}~~ a\in J_m, ~~ {\rm and}~~ x_{a,0} =\frac{\langle 1~a\rangle\langle 2~3\rangle }{\langle 1~3\rangle\langle 2~a\rangle}~~ {\rm for}~~ a\notin J_m.
\ee    
Here we have introduced the holomorphic spinor
\be 
\lambda_{I_m,\alpha} := \left(k_4+k_5+\cdots + k_m \right)_{\alpha,\dot\alpha}\tilde\mu^{\dot\alpha} 
\ee 
so that, for example, 
\be
\langle 1 ~ I_m\rangle = \langle 1|\left(k_4+k_5+\cdots + k_m \right)|\tilde\mu ].
\ee

The subleading order corrections are not illuminating, however, note that for $a,b\in J_m$, the leading order of $u_a-u_b$ cancels out. As it turns out, the $1/z$ term of $u_a-u_b$ for $a,b\in J_m$  is only slightly more complicated than the leading order and it is in fact very useful in the algorithm we introduce in section \ref{chAlg}. Here we present the expression
\be\label{subleadingDifference}
u_a-u_b = x_{a,1}-x_{b,1} + {\cal O}(z^{-2}) = -\frac{\langle 1~I_m\rangle\langle 3~I_m\rangle\langle a~b\rangle (k_4+k_5+\cdots +k_m)^2}{\langle 2~I_m\rangle\langle 1~3\rangle^2\langle a~I_m\rangle\langle b~I_m\rangle}\frac{1}{z}+ {\cal O}(z^{-2}). 
\ee 

The reader familiar with the CHY formulation should find the formula \eqref{leadingOrder} strangely familiar. In fact, if we allowed the set $J_m$ to contain a single element, then the formula in \eqref{leadingOrder} would become the exact MHV solution of the scattering equations! In a sense, the leading order solution in a general subsector has the structure of a MHV solution for a system with $n-m+1$ particles, i.e., $\{ 1,2,3, I_m,m+1,\ldots ,n\}$. Another curious and non-obvious fact is that the leading order of the solution in the subsector $J_n$ matches that of the $\overline{\rm MHV}$ solution\footnote{Here we take the Risager deformation applied to the standard $\overline{\rm MHV}$ solution to the scattering equations and keep only the leading order in the large $z$ limit.}. In a sense, the leading order of the solutions in the NMHV sector ``interpolate" between the MHV and $\overline{\rm MHV}$ solutions.

\section{Algorithm for Generating the Large $z$ Expansion and Radius of Convergence} \label{chAlg}

In this section we detail an algorithm for generating terms in the expansion $u_a = \sum_{i=0}^\infty x_{a,i}/z^i$. 
\vskip 0.2in
{\bf Algorithm 4.1:} (Expansion Around Infinity of NMHV Solutions)
\vskip 0.1in
{\bf Input:} $n$-point scattering equations with a Risager deformation of kinematics with parameter $z$ as in \eqref{risagerDeformation}. Integer $N$. Set $J_m\subset \{1,2,\ldots, n\}$.
\vskip 0.1in
{\bf Output:} Expansion of NMHV solutions up to order $N$, i.e. $u_a = \sum_{i=0}^N x_{a,i}/z^i + {\cal O}(1/z^{N+1})$ in the $J_m$ sector. 
\vskip 0.1in
Fix the gauge $u_1=0, u_2=\infty, u_3=1$. Initialize a list $C$ with the coefficient assignments $x_{a,i}= 0$ for $4\leq a\leq n$ and $1\leq i\leq N$ and \eqref{leadingOrder} for $x_{a,0}$. 

Expand the scattering equations $\frac{\partial {\cal S}(z)}{\partial u_a}$ to order $(1/z)^0$ for $a\in \{1,3\}\cup \overline{J}_m$ and to order $(1/z)^{-1}$ for $a\in J_m$. These are linear equations in $x_{a,1}$. The system has corank one and solving it gives rise to a one-parameter family of solutions. To completely fix the $x_{a,1}$'s, impose \eqref{subleadingDifference} for any pair $a,b\in J_m$.

Update $C$ with the values for $x_{a,1}$ with $4\leq a\leq n$.

The $k^{\rm th}$-order in the expansion is computed  recursively by substituting the $(k-1)^{\rm th}$-order expansion for the $u_a$'s and expanding the scattering equations to order $(1/z)^{(k-1)}$ for $a\in \{1,3\}\cup \overline{J}_m$ and to order $(1/z)^{(k-2)}$ for $a\in J_m$. Repeat this procedure until the $N^{\rm th}$ order is obtained and return $C$.

\vskip 0.2in

Using Algorithm 4.1 we have computed expansions for $n\leq 9$ with $N=300$.  

Having the ability of generating a large number of terms in the $1/z$ expansion of $u_a$, the natural question is its radius of convergence. As it turns out, this is more interesting than expected. 

The actual functions $u_a(z)$ are complicated functions with poles and branch points. The location of the poles is easy to find by noticing that the scattering equations mirror physical singularities in the moduli space ${\cal M}_{0,n}$. For example, solving the equation $s_{2a}(z)=0$ gives the value of $z=z_{2a}$ where $u_2 \sim u_a$ and since our gauge fixing has $u_2=\infty$, it means that for some of the solutions, $u_a(z)$ must have a simple pole at $z=z_{2a}$. Branch point singularities are harder to determine, as they are associated to points where two solutions coincide.

Since we have access to large orders in the expansion, one can compute the radius of convergence numerically and if it does not coincide with any of the poles, then it gives the location of a branch point. 

In our computations, we assume that the various kinematic invariants, $\langle a~b\rangle$ and $[a~b]$ are real and generically distinct. Note that this is not possible in Minkowski signature and requires $(2,2)$ signature. In this section we assume this to be the case only for simplicity. 

Using the algorithm decribed above, the series expansions, $u_a = \sum_{i=0}^\infty x_{a,i}/z^i$ for $a\in \{4,5,\ldots ,n\}$, have real coefficients $x_{a,i}$. The naive approach for computing the radius of convergence, i.e., as the limit of the sequence $r_i=x_{a,i}/x_{a,i+1}$ when $i\to \infty$ turns out to fail as the $r_i$ seem to have random behavior and the sequence does not converge. 

As it turns out, this is the generic behavior of real-valued functions with singularities at complex locations controlling the radius of convergence. The way to deal with this situation is to use the Mercer-Roberts estimator \cite{mercer1990centre}. We review the construction and give some examples in appendix \ref{MercerRoberts}. Here we simply point out that the procedure is motivated by noting that the series under consideration behaves as 
\be 
f(x) = \sum_{j=0}^\infty b_j \cos{(j\theta)} x^j
\ee 
for some $\theta \in \mathbb{R}$ which is controlled by the position of the singularities in the complex plane.

Note that while the radius of convergence is controlled by $b_j$, the sequence 
$$\left\{ \frac{b_j\cos{(j\theta)}}{b_{j+1}\cos{((j+1)\theta)}},~~ j\in \mathbb{N}  \right\} , $$ 
could seem random since $\cos{(j\theta)}$ is not periodic in $j$ for generic $\theta$ (in fact, $\cos{(j\theta)}$ is periodic only for $ \theta /\pi \in \mathbb{Q} $.). 

To extract the radius of convergence of our series, following Mercer and Roberts, one computes
\be
R_a := \lim_{j\to \infty} \frac{1}{B_{a,j}}, \quad {\rm with} \quad B_{a,j}^2 := \frac{x_{a,j+1}x_{a,j-1}-x_{a,j}^2}{x_{a,j}x_{a,j-2}-x_{a,j-1}^2}.
\ee 

Of course, in practice one only has access to a finite number of terms in the series. However, the plot of $\{1/j,B_{a,j}\}$ should have as the vertical intercept 
the reciprocal of the radius of convergence. Using the plot with a finite number of terms and a linear fit, easily provides the vertical intercept.

\section{NMHV Gravity Amplitudes}\label{chNMHV}

Tree level amplitudes of gravitons can be computed using BCFW recursion relations since the deformed amplitude vanishes as the deformation parameter is taken to infinity \cite{Benincasa:2007qj,Arkani-Hamed:2008bsc}. In fact, graviton amplitudes behave as ${\cal O}(z^{-2})$ as $z\to \infty$.

At first sight, a Risager deformation is not very different from a BCFW deformation. This made it surprising that NMHV $n$-graviton amplitudes behave as ${\cal O}(z^{n-12})$ as $z\to \infty$ and hence cannot be constructed from unitarity alone for $n\geq 12$ \cite{Bianchi:2008pu}. In fact, the proof of the ${\cal O}(z^{n-12})$ is cumbersome and not illuminating (see appendix B of \cite{Benincasa:2007qj}).  

In this section we use the CHY formulation of graviton amplitudes to provide a more direct proof of the ${\cal O}(z^{n-12})$ behavior and use the CSM formulation to compute the precise form of the leading order term for any $n$-graviton amplitude in an expansion around $z\to \infty$. 

\subsection{Derivation of Large $z$ Behavior \`a la CHY}

Before discussing gravity amplitudes, we start with gluon amplitudes. It is known that gluon amplitudes admit a MHV expansion and its derivation using a Risager deformation relies on the fact that there is no pole at infinity. Using the CHY formulation of gluon and graviton amplitudes we use the behavior at infinity of the former to derive that of the latter. In fact, the computation is so straightforward, that we obtain the behavior at infinity of each of the $2^{n-3}-n+2$ NMHV subsectors. 

Gluon amplitudes in the CHY formulation are computed using the solutions to the scattering equations and two matrices, traditionally denoted $\Phi(s_{ab},u_a)$ and $\Psi(\epsilon_a\cdot\epsilon_b,\epsilon_a\cdot k_b,s_{ab},u_a)$. Unfortunately, $\Phi$ is also the name of a different matrix which enters  in the CSM formulation. We hope that the context will prevent any confusions.  

We have made explicit the dependence of each matrix on the kinematic data and polarization vectors $\epsilon_a$ of the gluons. The explicit definition of $\Psi$ is not important for our purposes and can be found in \cite{Cachazo:2013hca}. The matrix $\Phi$ is a $n\times n$ symmetric matrix with components defined as
\be \label{phicom}
\Phi_{ab} := \frac{\partial^2 {\cal S}}{\partial u_a\partial u_b},
\ee
where ${\cal S}$ was given in \eqref{potentialCHY}.

It is easy to show that $\Phi$ has corank 3  when evaluated on any solution of the scattering equations. So, the reduced determinant is 
\be  
{\rm det}'\Phi := \frac{{\rm det}\Phi^{ijk}_{prq}}{|ijk||prq|},
\ee 
where $\Phi^{ijk}_{prq}$ is the matrix obtained from $\Phi$ by deleting the rows $i,j,k$ and the columns $p,q,r$ and $|abc|:=(u_a-u_b)(u_a-u_c)(u_b-u_c)$. The reduced determinant, ${\rm det}'\Phi$, is independent of the choices made for $i,j,k$ and $p,q,r$. 

The CHY formula for the partial amplitude of gluons is given by a sum over the solutions to the scattering equations,
\be\label{chyYM}  
A(1,2,\ldots ,n) = \sum_{\rm sols}\frac{{\rm Pfaff}'\Psi(\epsilon_a\cdot\epsilon_b,\epsilon_a\cdot k_b,s_{ab},u_a)}{{\rm det}'\Phi(s_{ab},u_a)}\frac{1}{(u_1-u_2)(u_2-u_3)\cdots (u_n-u_1)}.
\ee 
This formula is valid in any helicity sector. However, something remarkable happens when the sector defined by the polarization vectors and that of the solutions do not match. It turns out that the Pfaffian vanishes and hence we can compute the NMHV amplitudes using only the $k=3$ solutions to the scattering equations.  

Under the Risager deformation and using the MHV expansion of $A(1,2,\ldots ,n) $, one can show that $A(z)={\cal O}(1/z^4)$ as $z\to \infty$. This is strictly speaking an upper bound and to make it tight one would have to use, e.g., the Witten-RSV formulation.

Since we are interested in the leading order behavior of solutions in a given subsector, we should look for solutions that maximize the powers of $z$ due to the Parke-Taylor factor in \eqref{chyYM}. This happens precisely for the sets $J_m=\{4,5,\ldots ,m\}$ defined in section \ref{chSects}. Assuming we are looking at the solution in the $J_m$ sector, one can show that (see appendix B for a derivation)
\be\label{phiB} 
{\rm det}'\Phi(z) = {\cal O}(z^{n+m-7})~~{\rm as}~~z\to \infty .
\ee 
Combining this result with 
\be 
A(1,2,\ldots ,n)(z) ={\cal O}(z^{-4})\quad {\rm and}\quad \frac{1}{(u_1-u_2)(u_2-u_3)\cdots (u_n-u_1)}={\cal O}(z^{m-4}),
\ee 
one immediately concludes that in the subsector, $J_m$, under consideration
\be\label{psiB} 
{\rm Pfaff}'\Psi(\epsilon_a\cdot\epsilon_b,\epsilon_a\cdot k_b,s_{ab},u_a) = {\cal O}(z^{n-7}).
\ee 

Now we are ready to study the large $z$ behavior of gravity amplitudes. In their CHY formulation, gravity amplitudes are given by 
\be\label{chyGR}  
M_n= \sum_{\rm sols}\frac{\left({\rm Pfaff}'\Psi(\epsilon_a\cdot\epsilon_b,\epsilon_a\cdot k_b,s_{ab},u_a)\right)^2}{{\rm det}'\Phi(s_{ab},u_a)}.
\ee 
Combining \eqref{chyGR}, \eqref{phiB} and \eqref{psiB}, it is easy to obtain the result that the solution in the $J_m$ subsector gives a contribution to the amplitude of the form 
\be\label{largeZJm}  
M^{J_m~{\rm subsector}}_n(z) =  {\cal O}(z^{n-m-7})  \quad {\rm as}\quad z\to \infty.
\ee 

Unlike for gluon amplitudes, the large $z$ behavior of each subsector in the graviton amplitude depends on the value of $m$. This means that if one is interested in the overall leading order of the full graviton amplitude then one has to choose the value of $m$ that maximises $n-m-7$. Since $m\in \{5,6,\ldots ,n\}$, $m=5$ gives the maximum value and the leading behavior of the full amplitude is 
\be  
M_n(z) =  {\cal O}(z^{n-12})  \quad {\rm as}\quad z\to \infty ,
\ee 
as expected from previous results in the literature.

\subsection{The Twelve-Point Case: The Generalized MHV Expansion}

A twelve-graviton amplitude is the first case in which there is a pole at infinity that corrects the sum over MHV diagrams obtained from the poles when kinematic invariants vanish in the Risager construction applied to gravity. 

As explained above, NMHV solutions to the scattering equations give rise to contributions to the amplitude with different large $z$ behaviors. The most singular sector is that with $|J|=2$, i.e., with two labels in the set $J$. 

Using the Witten-RSV equations and the leading order solutions for the $u_a$'s found in section \ref{chSects}, one can use the CSM formula for the graviton amplitude in the NMHV sector with helicities, $M(1^{-},2^{-},3^{-},4^{+},\ldots ,n^{+})$, to get, after some algebra\footnote{The CSM formula is invariant under the choice of rows and columns that we eliminate from $\Phi$ and $\tilde{\Phi}$, but in the large $z$ limit, this is only manifest if we keep the leading and subleading order of the variables during the computations.}, the contribution to the amplitude of the $J=\{s_1,s_2\}$ sector 
 \be
         M^{(s_1,s_2)}_n(z) = -\frac{(-1)^{w_n}}{z^{12-n}}\frac{\langle s_1~s_2\rangle^3 [s_1~s_2]^7(\langle 1~2\rangle\langle 2~3\rangle\langle 3~1\rangle)^{n-6}}{[\tilde{\mu}~ s_1][\tilde{\mu}~ s_2]\prod_{i\in \{1,2,3,s_1,s_2\}}\langle i|s_1+s_2|\tilde{\mu}]} \prod_{j\notin \{1,2,3,s_1,s_2\}} \frac{[\tilde{\mu}~ j]}{\langle 1~j\rangle\langle 2~j\rangle\langle 3~j\rangle}, 
 \ee
where $w_n = \frac{1}{2}n(n+1)$. Note that the little group properties of $M^{(s_1,s_2)}_n(z)$ in particles $1,2$ and $3$ are not manifest since $z$ carries little group weight (see \eqref{risagerDeformation}).  

Specializing to $n=12$, the Risager deformation of $M_{12}(z)$ gives rise to the following generalized MHV expansion,
\be\label{residueAtInfinity}  
M_{12}(0) = \frac{1}{2\pi i}\oint \frac{dz}{z}M_{12}(z) = ({\rm MHV~Diagrams}) + \sum_{\{s_1,s_2\}\subset \{4,5,\ldots ,n\}}M^{(s_1,s_2)}_{12}.
\ee 
The terms in the sum over all two-element subsets of $\{4,5,\ldots ,n\}$ is the contribution from the residue at $z=\infty$. Note that for $n=12$, $M^{(s_1,s_2)}_{n}(z)$ is $z$ independent and in \eqref{residueAtInfinity} we have omitted it.  

The residue at infinity was first computed by Conde and Rajabi in \cite{Conde:2012ik} using a combination of BCFW and Risager techniques. Our formula, although different in form, perfectly agrees with the Conde-Rajabi formula. However, since the BCFW procedure makes two particles special, the Conde-Rajabi formula obscures the symmetries of the residue at infinity. It is our hope that our new presentation will inspire a ``diagrammatic" interpretation. We discuss some possible directions in section \ref{chDis}.

\subsection{General Subsector: Leading Order Contribution to the Amplitude}

As in the previous section, we can use the Witten-RSV equations and the leading-order solutions for the $u_a$’s together with the CSM formula to get the large $z$ behavior of the amplitude in the sector $J = \{s_1,...,s_p\}$. In this case, the leading contribution to the amplitude can be written as 
\bes\label{newLOgravity}
    M^{(s_1,..,s_p)}_n &=& -\frac{(-1)^{w_n}}{z^{10+p-n}} \frac{(k^2_{s_1...s_p})^{6}(\langle 2~3\rangle \langle 1~2 \rangle \langle 3~1\rangle)^{n-p-4}\widetilde{\det}(\phi)}{\prod_{i\in \{1,2,3\}}\langle i|k_{s_1...s_p}|\tilde{\mu}]} \prod_{j \in \hat{J}} \frac{[\tilde{\mu}~ j]}{\langle 1~j\rangle\langle 2~j\rangle\langle 3~j\rangle}, \nonumber \\&&
\ens
where $\hat{J} := \{4,...,n\}/J$, $k_{s_1...s_p}:= \sum_{i=1}^p k_{s_i}$ and $\phi$ is a $p \times p$ matrix with elements

\bes\label{newHodges}
    \phi_{ij} &=& \frac{[s_i~s_j]}{\langle s_i~s_j \rangle} \quad i \not = j, \\
    \phi_{ii} &=& -\sum_{\substack{j=1 \\ j\neq i}}^p \phi_{ij} \left(\frac{\langle s_j| k_{s_1...s_p}|\tilde{\mu}]}{\langle s_i| k_{s_1...s_p}|\tilde{\mu}]}\right)^2.
\ens
Finally, $\widetilde{\det}(\phi)$ is defined as
\bes
\widetilde{\det}(\phi) := \frac{(-1)^f_l \det(\phi^f_l)}{\langle s_f| k_{s_1...s_p}|\tilde{\mu}]^2\langle s_l| k_{s_1...s_p}|\tilde{\mu}]^2},
\ens
with $\phi^f_l$  the matrix obtained by removing any $f^{\rm th}$ row and $l^{\rm th}$ column from $\phi$.

\section{Computing Amplitudes at Infinity}\label{chCInf}

NMHV amplitudes of gluons or gravitons can be computed in many different ways and we have reviewed some of them in this work. Each technique has its pros and cons. For example, the BCFW recursion relations produce analytic formulas at the expense of introducing spurious poles. The Witten-RSV, CSM, or CHY formulations require numerically solving polynomial equations and evaluating an integrand on them. Very nicely, each solution gives rise to an object that satisfies many physical properties not manifest in the BCFW results such as decoupling identities in Yang-Mills. 

Our expansion around infinity opens up a third kind of approach. Starting with a Risager deformation, one can construct a rational function that has fewer poles than the deformed amplitude, $A(z)$. The poles at finite locations of $A(z)$ are known, they correspond to solutions of $\sum_{a,b\in F}s_{ab}(z)=0$, where $F\subset \{1,2,\ldots ,n\}$ is a subset that contains at least one and at most two of $\{1,2,3\}$, i.e. $1\leq |F\cap \{1,2,3\}|\leq 2$. Let us denote $P_F^2(z):=\sum_{a,b\in F}s_{ab}(z)$. 

Now, supposed that we want to eliminate poles located at the solutions of $P_{F_1}^2(z)=0$, $P_{F_2}^2(z)=0$, $\ldots$, $P_{F_q}^2(z)=0$. The simplest way to achieve this is by considering the relation
\be\label{lessPoles}
A(0) = \frac{1}{2\pi i}\oint_{|z|=\epsilon} \frac{dz}{z}A(z)\prod_{j=1}^q \frac{P^2_{F_j}(z)}{P^2_{F_j}(0)}.
\ee 
The residue theorem expresses the integral as a sum over the residues of poles of the integrand, excluding $z=0$, and possibly including  $z=\infty$. 

Clearly, if the amplitude $A(z)/z$ behaves as ${\cal O}(z^{t-1})$ as $z\to \infty$ then the presence of the factors $P^2_{F_j}(z)$, which are linear in $z$, gives ${\cal O}(z^{t+q-1})$. If $t+q>0$ there is a pole of order $t+q+1$ at infinity. 

Now, the idea is to use any of the Witten-RSV, CSM, or CHY formulas and the algorithm described in section \ref{chAlg}, to compute the necessary residues at infinity. 

The most extreme use of \eqref{lessPoles} is when we eliminate all physical poles of the amplitude $A(z)$. In this extreme case 
\be
A(0) = -\frac{1}{2\pi i}\oint_{|z|=\Lambda \gg 1} \frac{dz}{z}A(z)\prod_{{\rm all}~j} \frac{P^2_{F_j}(z)}{P^2_{F_j}(0)}.
\ee
Assuming that the number of physical poles of $A(z)$ is $N_p$, one uses the expansions 
\be \label{eqExp}
A(z) = \sum_{i=-t}^\infty A_i \frac{1}{z^i} \quad {\rm and}\quad \prod_{{\rm all}~j} \frac{P^2_{F_j}(z)}{P^2_{F_j}(0)} = \sum_{i=-N_p}^0 G_i \frac{1}{z^i},
\ee 
to write what we call the {\it Formulation at Infinity} of an amplitude $A$,
\be
A(0) = \, -\!\!\!\!\!\!\!\!\!\sum_{i=\max(0,-t)}^{N_p}\!\!\!\!\!\!\! A_{i}\, G_{-i}.
\ee 

Note that this expansion is meant to be used as a (semi-)numerical technique. It would be interesting to study the efficiency of this technique compared to other numerical procedures. We expect this technique to be most useful in graviton amplitude computations. Nevertheless, we provide some examples involving gluon amplitudes in appendix \ref{appExamples}.

\section{Discussions}\label{chDis}

In this work we started the exploration of solutions to the scattering equations in regions of kinematic invariants around ``infinity". Here infinity refers to large values of the Risager deformation parameter for NMHV amplitudes. 

One of the very first results regarding solutions of the scattering equations was the fact that near physical singularities where the sum of momenta of a subset of particles, say $J\subset [n]$, becomes a null vector, some solutions become singular. If $|J|=m$, the $(m+1-3)!\times (n-m+1-3)!$ solutions become singular, i.e., correspond to points in the boundary of $\overline{{\cal M}}_{0,n}$. Such points have the description of two $\mathbb{CP}^1$, one with $m+1$ punctures and the other with $n-m+1$ punctures. The remaining solutions making up the rest of the $(n-3)!$ solutions remain at generic points in ${\cal M}_{0,n}$.

The behavior in the large $z$ limit of the Risager deformation is somewhat different since all solutions in the $k=3$ sector become singular. Moreover, unlike the factorization behavior, in which the singular solutions only split into the ``left" and ``right" sectors, in our case each solution becomes its own subsector. This is what allowed us to find the analytic form of the leading order solutions for all multiplicity. 

Another surprise was the fact that the leading order solutions have an MHV or $k=2$-like structure that seems to interpolate between the actual $k=2$ solutions and the $k=n-2$ solution. 

The most important next step is to explore other sectors, in particular the N$^2$MHV or $k=4$ sector. Here the Risager deformation is not as canonical as for the $k=3$ sector as it comes with more freedom. A preliminary analysis shows that the $E(n-3,2)=3^{n-3}-2^{n-3}(n-2)+(n-3)(n-2)/2$ solutions in the $k=4$ sector split into some subsectors but not completely as it happened for $k=3$. This might be related to the non-canonical nature of the deformation. 

Of course, it would be ideal to classify all deformations, even beyond the Risager type, that exhibit the complete splitting of sectors into subsectors as Risager's does for the NMHV sector.

Finally, we used the expansion to understand the large $z$ behavior of NMHV gravity amplitudes under the Risager deformation. The CHY formulation makes it clear that the ${\rm Pfaff}'\Psi$ integrand is the culprit for the bad behavior. It is very interesting that its large $z$ dependence on $n$ is completely cancelled by the Hessian of the potential ${\cal S}$, defined in \eqref{potentialCHY}, leaving Yang-Mills amplitudes with a $n$-independent large $z$ behavior. Once again, it would be interesting to repeat this analysis for other sectors. 

The $n$-dependent large $z$ behavior of NMHV amplitude only affects the residue theorem argument to compute the amplitude for $n\geq 12$ gravitons. We computed the residue at infinity of the twelve-graviton NMHV amplitude using the new expansions and found an expression that is a sum over all possible pairs of positive helicity gravitons. It is tempting to conjecture that either these terms can be given a novel diagrammatic interpretation or that they can be written in a form that ``corrects" each of the MHV or CSW diagrams produced by the Risager construction. Let us briefly discuss each possibility. 

In the large $z$ limit, one can think that three negative helicity ``hard" particles $1,2,3$ scatter in the presence of a background of ``soft" positive helicity gravitons. This is the same picture as the one used by Arkani-Hamed and Kaplan to study the behavior of amplitudes in the large $z$ limit under the BCFW deformation \cite{Arkani-Hamed:2008bsc}. From this point of view, one can try and rewrite the leading order contributions of each subsector in the large $z$ limit presented in eq. \ref{newLOgravity} in a form closer to this interpretation,
\bes\label{newLOgravity}
    M^{(s_1,..,s_p)}_n(z) &=& \frac{1}{w^{4}} \widetilde{\det}(\phi)\frac{(P^2)^{6}}{\prod_{m\in \{1,2,3\}}(P+k_m(z))^2} \left(\prod_{j \in \hat{J}} S_j(z)\right) M(1^-,2^-,3^-)+\ldots, \nonumber \\&&
\ens
where the ellipses represent lower orders in the $1/z$ expansion, $P\! =\! k_{s_1}+k_{s_2}+\ldots +k_{s_p}$, $\hat J = [n]\setminus\{ 1,2,3,s_1,s_2,\ldots ,s_p\}$, $w$ is a little group invariant version of $z$, i.e.
\be
w^2 = \langle 1~2\rangle\langle 2~3\rangle\langle 1~3\rangle\, z^2,
\ee
$S_j(z)$ is the leading soft factor for graviton $j$ coming off the  amplitude $M(1^-,2^-,3^-)$, i.e.
\be 
S_i(z) = \sum_{m=1}^3\frac{[i~\tilde\lambda_m(z)]\langle m~x\rangle\langle m~y\rangle}{\langle i~m\rangle\langle i~x\rangle\langle i~y\rangle}.
\ee 
and the three-point amplitude
\be\label{r2} 
M(1^-,2^-,3^-)= \left(\langle 1~2\rangle\langle 2~3\rangle\langle 1~3\rangle\right)^2,
\ee 
which is fixed by little group and non-singular behavior for real kinematics \cite{Benincasa:2007xk}. At tree-level Einstein gravity does not have three-point vertices that can produce \eqref{r2}. However, it is known that an $R^3$ term does. These kind of terms is generated by the well-known two-loop UV divergence in Einstein gravity. It would be interesting to understand the connection between the background computation and the $R^3$ term.

As for the second possible interpretation, already in Witten's original work \cite{Witten:2003nn}, it was pointed out that MHV graviton amplitudes are not localized on lines in twistor space but they have a ``derivative of a delta function" support. It could be that this failure to be localized on a line is an effect that cancels out for a small number of gravitons, i.e., $n<12$, and that for $n\geq 12$ is gives rise to the expression we found for the residue at infinity. Along this line of thought, one has to rewrite the leading order contributions from each subsector, \eqref{newLOgravity}, as objects as close as possible to MHV amplitudes. In the orginal CSW construction, one introduces an internal particle $I$ and writes NMHV amplitudes as sums over products of MHV amplitudes where the internal particle appears in both. Here we write
\be 
P_{\alpha \dot\alpha} = \frac{P^2}{\langle I~\mu\rangle[I~\tilde\mu]}\lambda_{I,\alpha}\tilde\lambda_{I,\dot\alpha}  + \mu_{\alpha}\tilde\mu_{\dot\alpha}.
\ee 
and treat $\mu_{\alpha}$ and $\tilde\mu_{\dot\alpha}$ as reference spinors. In fact, we take the latter to coincide with the reference spinor in Risager's construction. Using this definition one can write \eqref{newLOgravity} as
\bes\label{newLOgravityV2}
    M^{(s_1,..,s_p)}_n(z) &=& \frac{1}{w^{8}}\left( \frac{\langle I~\mu\rangle^{7}}{[I~\tilde\mu ]} \right) \widetilde{\det}(\phi(\lambda_I)) \left(\prod_{j \in \hat{J}\cup \{ I \}} S_j(z)\right) M(1^-,2^-,3^-)+\ldots, \nonumber \\&&
\ens
Note that in this formula the product is not only over particles in $\hat J$ but it also includes $I$ which we take to be defined by spinor $\lambda_{I,\alpha}$ and $\tilde\lambda_{I,\dot\alpha}$. Also, $\widetilde{\det}(\phi(\lambda_I))$ is slightly different from is original definition in \eqref{newHodges}. The matrix $\phi$ is still of size $p\times p$, the off-diagonal terms stay the same, while the diagonal ones become
\bes\label{newHodgesV2} 
    \phi_{ii} = -\sum_{\substack{j=1 \\ j\neq i}}^p \phi_{ij} \left(\frac{\langle s_j~I\rangle}{\langle s_i~I\rangle}\right)^2.
\ens
Finally, $\widetilde{\det}(\phi(\lambda_I))$ is now given by
\bes
\widetilde{\det}(\phi(\lambda_I)) := \frac{(-1)^f_l \det(\phi^f_l)}{\langle s_f~I\rangle^2\langle s_l~I\rangle^2}.
\ens

We leave a more detailed study of these formulas for future work.

\section*{Acknowledgements}

The authors thank Yong Zhang for discussions. This research was supported in part by a grant from the Gluskin Sheff/Onex Freeman Dyson Chair in Theoretical Physics and by Perimeter Institute. The research of PL was supported in part by ANID/ POSTDOCTORADO BECAS CHILE/ 2022 - 74220031. Research at Perimeter Institute is supported in part by the Government of Canada through the Department of Innovation, Science and Economic Development Canada and by the Province of Ontario through the Ministry of Colleges and Universities.

\appendix

\section{Radius of Convergence of Taylor Series of Real Functions with Singularities in the Complex Plane}\label{MercerRoberts}

When a real function $f(x)$ has a Taylor expansion around $x=0$ with finite radius of convergence $R$, the value of $R$ is determined by the location of the singularities of $f(z)$ as a function of a complex variable $z$. $f(z)$ can have many singularities but only the ones closest to the origin matter and their modulus equals $R$. When the only singularities with modulus $R$ are on the real axis, i.e., at $z=R$ or at $z=-R$, the standard procedure for finding $R$ from the coefficients $a_i$'s of 
\be
f(z) = \sum_{i=0}^\infty a_i z^i
\ee
is very simple. One computes
\be
R = \lim_{i\to \infty} \frac{a_i}{a_{i+1}}.
\ee 
Of course, if only a finite number of the coefficients is known, then one can plot $(1/i,a_i/a_{i+1})$ and use a linear interpolation to get the value of the vertical intercept.

As mentioned in section \ref{chAlg}, the expansion of the solutions to the scattering equations can have two singularities located off the real axis, i.e., at $z=R e^{i\theta}$ and at $z=R e^{-i\theta}$ for some generic argument $\theta$. When this is the case the sequence $\{a_i/a_{i+1}\}$ can exhibit some random behavior and $R$ cannot be determined in the standard way. 

In appendix B of \cite{mercer1990centre}, Merces and Roberts modeled the situation by using a simple function with only two singularities at the desired values. Their ``template" function was
\be 
g(z) = \left(1-\frac{z}{r e^{i\theta}}\right)^\nu + \left(1-\frac{z}{r e^{-i\theta}}\right)^\nu,
\ee 
where $\nu$ is not a positive integer.

Expanding around $z=0$, the series can be written as 
\be
g(z) = \sum_{i=0}^\infty A_i B^i \cos (i \theta) z^i,
\ee 
where $A_i$ is a function of $i$ that varies very slowly with $i$ for $i\gg 1$.

If one only has knowledge of the coefficients $a_i=A_i B^i \cos (i\theta)$, then $B$ can be extracted by using that
\be  
B^2 \sim \frac{a_{i+1}a_{i-1}-a_i^2}{a_{i}a_{i-2}-a_{i-1}^2}.
\ee 
This can be proven by assuming that $A_i\sim A_{i+1}\sim A_{i-1}\sim A_{i-2}$ and some trigonometric identities. 

Based on this, Mercer and Roberts propose that for any series where one suspects that there are exactly two singularities at $|z|=|R|$ and off the real axis, one can construct the sequence 
\be 
B_i^2 := \frac{a_{i+1}a_{i-1}-a_i^2}{a_{i}a_{i-2}-a_{i-1}^2}.
\ee 
and find its limit as $i\to \infty$ to get $1/R$. Once again, in practice, one plots the list $\{1/i,B_i\}$ and uses a linear fit to get the vertical intercept. 

\subsection{Example: $n=6$ Case}

We now employ the algorithm detailed in section \ref{chAlg}, to generate power series solutions of the scattering equations \eqref{rationalCHY}. That is, we obtain the coefficients of the power series $u_a(z) = \sum_{i=0}^N x_{a,i} / z^i$ and $N=300$. As discussed in section \ref{chSects}, after employing the gauge fixing $\{ u_1 =0, u_2=\infty, u_3=1\}$, the solutions separate into subsectors labeled by subsets $J \subset \{4,5,6\}$ with $|J|>1$. There are only four such subsectors given by $J^{(1)} = \{4,5\}, J^{(2)} = \{4,6\}, J^{(3)} = \{5,6\}, J^{(4)} = \{4,5,6\}$. Below we present the large order behavior of the coefficients of the power series expansion of $u_4(z)$ for the sectors $J^{(1)}$ and $J^{(4)}$ where we have used the kinematic data presented in table \ref{tab:kinematics}. Additionally, we chose our reference spinor to be $\tilde \mu_{\dot 1} = 2/193,\> \tilde \mu_{\dot 2} = 3/217$. 

\begin{table}[]
\caption{Kinematic data used in numerical evaluation of the scattering equations.}
\label{tab:kinematics}
\scalebox{1.0}{
\begin{tabular}{|l|l|l|l|l|l|l|l|l|l|l|l|l|}
\hline
\multirow{1}{*}{$a$} &
\multicolumn{2}{l|}{1} &
\multicolumn{2}{l|}{2} &
\multicolumn{2}{l|}{3} &
\multicolumn{2}{l|}{4} &
\multicolumn{2}{l|}{5} &
\multicolumn{2}{l|}{6} \\
\hline
$\alpha / \dot \alpha$ & 1 & 2 & 1 & 2 & 1 & 2 & 1 & 2 & 1 & 2 & 1 & 2 \\
\hline
$\lambda_{a,\alpha}$ & 47 & 61 & 103 & 113 & 167 & 181 & 233 & 251 & 307 & 317 & 379 & 397 \\
\hline
$\tilde \lambda_{a,\dot \alpha}$ & 53 & 97 & 83 & 131 & 113 & 167 & 157 & 199 & $\frac{267391}{434}$ & $\frac{398327}{434}$ & $-\frac{292735}{434}$ & $-\frac{428359}{434}$\\
\hline
\end{tabular}
}
\end{table}

\begin{figure}%
    \centering
    \subfloat[\centering]{{\includegraphics[width=7cm]{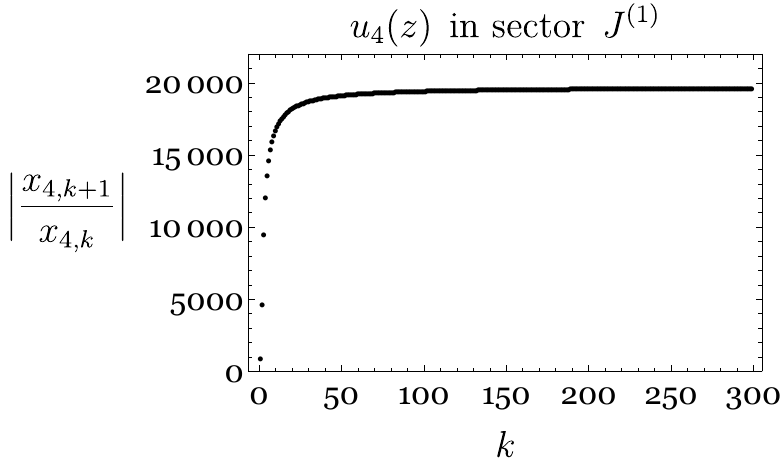} }}%
    \qquad
    \subfloat[\centering]{{\includegraphics[width=7cm]{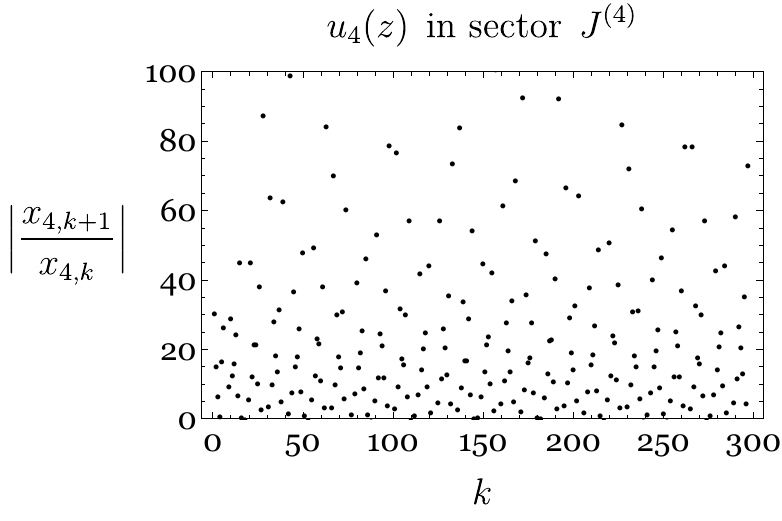}}}%
    \caption{Ratio of coefficients: (a) $u_4(z)$ in the sector $J_1$  (b) $u_4(z)$ in the sector $J_4$.}%
    \label{fig:regconv}%
\end{figure}

For the sector $J^{(1)}$ the ratio test is sufficient for estimating the radius of convergence of the power series expansion as shown in figure \ref{fig:regconv} (a). We estimate the inverse of the radius of convergence by performing a linear fit of the data $(1/k, |x_{4,k+1}/x_{4,k}|)$ and extracting the vertical intercept. We find that $R^{-1} = 20,432.8$ for sector $J^{(1)}$. We expect that this radius of convergence is controlled by the location of the singularity of $u_4(z)$ nearest to the origin. Indeed, for sector $J^{(1)}$, we find that the singularity of $u_4(z)$ nearest to the origin is a branch point located at $|z_{*}|^{-1} = 20,432.8$.

\begin{figure}%
    \centering
    \subfloat[\centering ]{{\includegraphics[width=7cm]{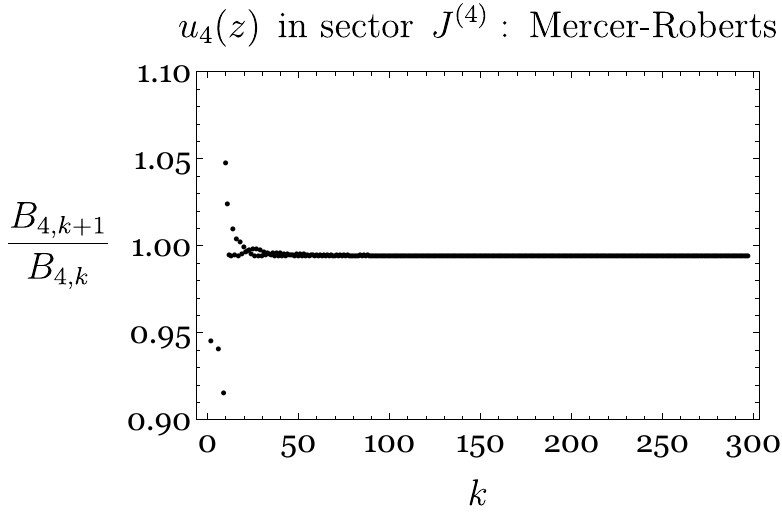} }}%
    \qquad
    \subfloat[\centering]{{\includegraphics[width=7cm]{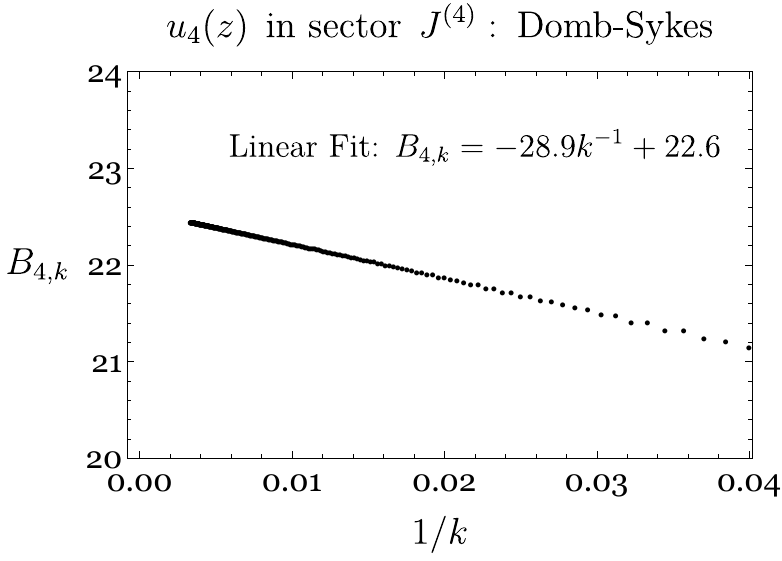}}}%
    \caption{(a) Mercer-Roberts coefficients for $u_4(z)$ in the sector $J_4$. (b) Domb-Sykes intercept plot for $u_4(z)$ in the sector $J_4$.}%
    \label{fig:MRDS}%
\end{figure}

On the other hand, as figure \ref{fig:regconv} (b) shows, for sector $J^{(4)}$, the ratio of consecutive terms displays a random behavior. However, the corresponding Mercer-Roberts sequence shown in figure \ref{fig:MRDS} (a) suggests that the original power series is convergent. The inverse of the radius of convergence is estimated by the vertical intercept of the Domb-Sykes plot shown in figure \ref{fig:MRDS} (b). We find that $R^{-1} = 22.6$. We also find that $u_4(z)$ has branch points at a pair of complex conjugate locations with modulus $|z_*|^{-1} = |\bar z_*|^{-1} = 22.6$.

%


\section{Derivation of Large $z$ Behavior of ${\rm det}'\Phi$}

Let us begin by analyzing the large $z$ behavior of the $\Phi$ components. From (\ref{phicom}), it is easy to check that, for $a,b\in \{4,...,m\}$ 

\bes \label{com1}
    \Phi_{ab} &=&  z^2\frac{\langle a ~ b \rangle [a~b]}{(x_{a,1}-x_{b,1})^2} + {\cal O}(z), \quad a\neq b \\ \Phi_{aa} &=& -z^2 \sum_{\substack{c=4 \\ c\neq a}}^m \frac{\langle a ~ c \rangle [a~c]}{(x_{a,1}-x_{c,1})^2} + {\cal O}(z),
\ens
while for $a\in \{4,...,m\}$ and $b \in \{m+1,...,n\}$ 

\bes \label{com2}
    \Phi_{ab} &=& \frac{\langle a ~ b \rangle [a~b]}{(x_{a,0}-x_{b,0})^2} + {\cal O}(1/z).
\ens
For $a,b \in \{m+1,...,n\}$

\bes \label{com3}
  \Phi_{ab} &=& \frac{\langle a ~ b \rangle [a~b]}{(x_{a,0}-x_{b,0})^2} + {\cal O}(1/z), \quad a\neq b \\ \Phi_{aa} &=& - z\sum_{c=1}^3 \frac{\langle a ~ c \rangle  L_c [\tilde{\mu}~c]}{(x_{a,0}-u_{c,0})^2} + {\cal O}(1),
\ens
where $L_a=\langle b~c\rangle$ and  $(a, b, c)$ is a cyclic permutation of $(1, 2, 3)$. The remaining cases are $a \in \{1,2,3\}$ and $b \in \{4,...,n\}$ in which 
\bes \label{com4}
\Phi_{ab} &=& z \frac{L_a \langle a~b\rangle [\tilde{\mu}~b]}{(u_a - x_{b,0})^2} + {\cal O}(1), 
\ens
 and, finally, for $a,b \in \{1,2,3\}$

\bes \label{com5}
\Phi_{ab} &=& z \frac{\langle a~b\rangle (L_a  [\tilde{\mu}~b] - L_b [\tilde{\mu}~a])}{(u_a - u_b)^2} + {\cal O}(1). \\
\Phi_{aa} &=& -z \left(\sum^3_{\substack{c=1 \\ c \neq a}} \frac{\langle a~c\rangle (L_a  [\tilde{\mu}~c] - L_c [\tilde{\mu}~a])}{(u_a - u_c)^2}+\sum^n_{c=4}\frac{L_a \langle a~c\rangle [\tilde{\mu}~c]}{(u_a - x_{c,0})^2}\right) + {\cal O}(1).
\ens
In order to derive the large $z$ behavior of ${\rm det}'\Phi$ we chose to eliminate the rows and columns $1,2,3$ of $\Phi$. Now, defining

\bes
A := \left(\begin{array}{ccc}
     \frac{\Phi_{44}}{z^2} & \cdots &  \frac{\Phi_{4m}}{z^2} \\
     \vdots & \ddots & \vdots \\
     \frac{\Phi_{m4}}{z^2} & \cdots &  \frac{\Phi_{mm}}{z^2}
\end{array}\right), ~~  B := \left(\begin{array}{ccc}
     \frac{\Phi_{4 m+1}}{z} & \cdots &  \frac{\Phi_{4 n}}{z} \\
     \vdots & \ddots & \vdots \\
     \frac{\Phi_{n m+1}}{z} & \cdots &  \frac{\Phi_{m n}}{z}
\end{array}\right), ~~  C := \left(\begin{array}{ccc}
     \frac{\Phi_{m+1 m+1}}{z} & \cdots &  \frac{\Phi_{m+1 n}}{z} \\
     \vdots & \ddots & \vdots \\
     \frac{\Phi_{n m+1}}{z} & \cdots &  \frac{\Phi_{n n}}{z}
\end{array}\right)     
\ens
we can write
\bes
\det(\Phi_{123}^{123}) &=& z^{n+m-6}\det(C)\det\left(A-\frac{1}{z}B~C^{-1}B^T\right)\nonumber \\ &=& z^{n+m-6}(\det(C)\det(A))_{z \rightarrow \infty} + {\cal O}(z^{n+m-7}).   
\ens
Finally, using (\ref{com1}-\ref{com5}), it is easy to check that $(\det(A))_{z \rightarrow \infty} = 0$ since the rows of $A$ add up to zero in the large $z$ limit, and therefore
\bes
\det(\Phi_{123}^{123}) =  {\cal O}(z^{n+m-7}).  
\ens

\section{Computing Amplitudes at Infinity: Examples}
\label{appExamples}

In this appendix we provide examples of how to use the expansions around $z\to \infty$ for the computation of gluon amplitudes. While the technique is expected to be most useful in graviton amplitude computations, we discuss gluon amplitudes as an illustration. 

The starting point of our computation is the Witten-RSV formula written in the canonical gauge $\{t_1 = 1, u_1 = 0, u_2=\infty, u_3=1\}$. Before we gauge fix the integral \eqref{eqWRSV}, we perform the change of variables $t_2 \rightarrow \tau_2 / u_2^d$ which is designed to make the limit $u_2 \rightarrow \infty$ of the integral manifestly finite. Now, we can freely gauge fix the integral to canonical gauge. This will introduce the Jacobian:
\be
J = t_1 (u_1 - u_2)(u_2 - u_3)(u_3 - u_1).
\ee
The measure in canonical gauge, then, reads:
\be
d \mathcal{M}_{n,d}^c = \prod_{m=0}^d d^{2}\rho_m \> \frac{d \tau_2}{\tau_2} \prod_{a=3}^n \frac{d t_a}{t_a}\prod_{a=4}^n d u_a \Big[\frac{1}{(1- u_4)(u_4 - u_5) \cdots (u_{n-1} - u_n)u_n}\Big].
\ee

After evaluating the delta functions on the canonical gauge, we obtain the following gauge-fixed integral:

\begin{multline} \label{canWRSV}
A_n^{(k=d+1)} = \int d \mathcal{M}_{n,d}^c \>\> \delta^2(\lambda_1 -\rho_0)\delta^2(\lambda_2 - \tau_2 \rho_d)\delta^2(\lambda_3 - t_3\sum_{m=0}^d \rho_m)\prod_{a=4}^n \delta^2(\lambda_a - t_a \sum_{m=0}^d \rho_m u_a^m)\\ \times \prod_{m=0}^d \delta^{2|4}(\delta_{m,0} \tilde \Lambda_1 + \delta_{m,d} \tau_2 \tilde \Lambda_2 + t_3 \tilde \Lambda_3 + \sum_{a=4}^n t_a u_a^m \tilde \Lambda_a),
\end{multline}
where we package the anti-holomorphic spinors and four-component Grassmann variables as $\tilde \Lambda_a = (\tilde \lambda_a, \eta_a)$.

Now we specialize to the case of interest, i.e., $A(1^-,2^-,3^-,4^+,\ldots ,n^+)$. This is done by setting $k=3$, $d=2$, and replacing the Grassmann variable delta functions by $(\tau_2 t_3)^4$. 

Moreover, we restrict our analysis to $n=6$ so that the amplitude is simple enough to allow us to carry out some analytic computations and compare them to the numerical technique we want to illustrate. 

Let us start with the BCFW formula for the six-point amplitude \cite{Britto:2004ap},
\be
A(1^-,2^-,3^-,4^+,5^+,6^+) = \frac{\lgg 3|1+2|6]^3}{P_{126}^2[21][16]\lgg 34 \rgg\lgg 45\rgg\lgg 5|1+6|2]} + \frac{\lgg 1|5+6|4]^3}{P_{156}^2[23][34]\lgg 56\rgg\lgg 61\rgg\lgg 5|1+6|2]}.
\ee

Applying Risager deformation \eqref{risagerDeformation} to this amplitude and expanding around $z=\infty$ allows us to compute the leading coefficients of the amplitude $A(z) = A_4 1/z^4 + A_5 1/z^5 + A_6 1/z^6 + \mathcal{O}(1/z^7)$. For instance, the leading term $A_4$ is given explicitly by:
\begin{multline}
A_4 = \frac{1}{\lgg 21 \rgg \lgg 23 \rgg (\lgg 51 \rgg \lgg 3| 2 |\tilde \mu] + \lgg 3,1 \rgg \lgg 5| 1 + 6 |\tilde \mu] )} \Big[ \frac{\lgg 3|1 + 2|6]^3}{\lgg 34 \rgg \lgg 45 \rgg \lgg 3|1+2|\tilde \mu] \lgg 3|4+5|\tilde \mu] [6\tilde \mu]}\\
+ \frac{\lgg 1|5 + 6|4]^3}{\lgg 56 \rgg \lgg 61 \rgg \lgg 1|2+3|\tilde \mu] \lgg 1|5+6|\tilde \mu] [\tilde \mu 4]} \Big].    
\end{multline}
The higher order terms $A_5,A_6$ can be computed in a similar fashion. We have reproduced these coefficients numerically by applying the subsector solutions of the scattering equations around $z=\infty$ to the canonically gauge fixed Witten-RSV integral \eqref{canWRSV}.\\

We end this appendix with an example for the six-point gluon amplitude.  

It is easy to check that under the Risager deformation, $A(z)\to 1/z^4$ as $z\to \infty$. The color ordering restricts the physical poles to have the form $1/s_{i,i+1}$, $1/s_{i,i+1,i+2}$. Of these, only the ones that are $z$ dependant are relevant. These are $1/s_{12}(z)$, $1/s_{23}(z)$, $1/s_{34}(z)$, $1/s_{61}(z)$,$1/s_{234}(z)$, $1/s_{345}(z)$.

We choose to eliminate all collinear poles, thus producing a generalized CSW formula where only MHV-diagrams corresponding to multiparticle singularities contribute.

We eliminate $1/s_{12}(z)$, $1/s_{23}(z)$, $1/s_{34}(z)$, and $1/s_{61}(z)$. This means that $N_p=4$. Noting that in this case $t=-4$ one has
\be
A(0) = (C_{234} + C_{345}) - A_4 G_{-4},
\ee 
where $C_{234},C_{345}$ are terms of the modified CSW expansion associated with the poles $1/s_{234}(z)$, $1/s_{345}(z)$ and $G_{-4}$ is the $z^4$ coefficient, i.e., leading coefficient of the singularities we want to eliminate, $R(z) = s_{12}(z)s_{23}(z)s_{34}(z)s_{61}(z)/s_{12}(0)s_{23}(0)s_{34}(0)s_{61}(0)$.

The modified CSW terms are given by:
\bes
C_{234} &= A^{\text{MHV}}(2^-, 3^-, 4^+, I^+_{234})\frac{R(z_{234})}{P_{234}^2}A^{\text{MHV}}(I_{234}^-, 5^+, 6^+, 1^-)\\
C_{345} &= A^{\text{MHV}}(3^-, 4^+, 5^+, I^-_{345})\frac{R(z_{345})}{P_{345}^2}A^{\text{MHV}}(I_{345}^+, 6^+, 1^-, 2^-),
\ens

where $I_{234}$ has the holomorphic spinor $\lambda_{I_{234}} = P_{234}|\tilde \mu]$ and $I_{345}$ has the holomorphic spinor  $\lambda_{I_{345}} = P_{345}|\tilde \mu]$. The locations $z_{234}$ and $z_{345}$ are the zeros of $s_{234}(z), s_{345}(z)$ respectively. They are given by the expressions:

\be
z_{234} = \frac{s_{234}(0)}{\lgg 3 1 \rgg \lgg 2 I_{234}\rgg + \lgg 12 \rgg \lgg 3I_{234}\rgg}, \>\> z_{345} = \frac{s_{345}(0)}{\lgg 12 \rgg \lgg 3I_{345} \rgg}.
\ee

We have also checked that the extreme case, i.e., give the Formulation at Infinity of $A(1^-,2^-,3^-,4^+,5^+,6^+)$. Now $N_p=6$ and
\be
A(0) = - \left(A_4 G_{-4}+A_5 G_{-5}+A_6 G_{-6}\right)
\ee 
agrees with the expected result.

\bibliographystyle{JHEP}
\bibliography{references}

\end{document}